\documentclass{raa}          
\usepackage{graphicx,times}
\usepackage{natbib}
\usepackage{amssymb,amsmath}
\bibpunct{(}{)}{;}{a}{}{,}

\usepackage[a4paper=true,pagebackref=true]{hyperref}
\hypersetup{pdftitle = The title of my PDF, pdfauthor = My name, pdfsubject= The subject, pdfkeywords = keyword1 keyword2 keyword3} 
\hypersetup{colorlinks = true, linkcolor = green, anchorcolor = red, citecolor = blue, filecolor = red, pagecolor = red, urlcolor = red}

\begin{document}

\title{Binary Star Fractions from the LAMOST DR4}

 \volnopage{ {\bf 2012} Vol.\ {\bf X} No. {\bf XX}, 000--000}
   \setcounter{page}{1}

   \author{Zhijia Tian\inst{1}$^*$, Xiaowei Liu\inst{1,4}, Haibo Yuan\inst{2}, Bingqiu Chen\inst{4}$^*$, Maosheng Xiang\inst{2}$^*$, Yang Huang\inst{1,4}$^*$,
   Chun Wang\inst{1}, Huawei Zhang\inst{1}, Jincheng Guo\inst{1}$^*$, Juanjuan Ren\inst{3}, Zhiying Huo\inst{3}, Yong Yang\inst{4}, 
   Meng Zhang\inst{1}, Shaolan Bi\inst{2}, Wuming Yang\inst{2}, Kang Liu\inst{2}, Xianfei Zhang\inst{2}, Tanda Li\inst{3}, 
   Yaqian Wu\inst{2}, Jinghua Zhang\inst{2}
   \footnotetext{\small $*$ LAMOST Fellow.}
   }

   \institute{ Department of Astronomy, Peking University, Beijing 100871, China; {\it tianzhijia@pku.edu.cn; x.liu@pku.edu.cn}\\
        \and
             Department of Astronomy, Beijing Normal University, Beijing 100875,  P.R.China\\
        	\and
            National Astronomy Observatories, Chinese Academy of Sciences, Beijing 100012, P.R.China\\
        \and 
            Department of Astronomy, Yunnan University, Kunming 650091,  P.R.China\\
\vs \no
   {\small Received 2012 June 12; accepted 2012 July 27}
}

\abstract{Stellar systems composed of single, double, triple or high-order systems are rightfully regarded as the fundamental building blocks of the Milky Way.  
Binary stars play an important role in  formation and evolution of the Galaxy. 
Through comparing the radial velocity variations from multi-epoch observations, we analyze the binary fraction of dwarf stars observed with the LAMOST. 
Effects of different model assumptions such as orbital period distributions on the estimate of binary fractions, are investigated. 
The results based on log-normal distribution of orbital periods reproduce the previous complete analyses better than the power-law distribution. 
We find that the binary fraction increases with $T_{\rm eff}$ and decreases with [Fe/H]. 
We first investigate the relation between $\alpha$-elements and binary fraction in such a large sample as the LAMOST. 
The old stars with high [$\alpha$/Fe]  dominate higher binary fraction than young stars with low [$\alpha$/Fe]. 
At the same mass, former forming stars possess a higher binary fraction than newly forming ones, which may be related with the evolution of the Galaxy.
}

   \authorrunning{Tian et al. }            
   \titlerunning{Binary Star Fractions from the LAMOST DR4}  
   \maketitle

%
\section{Introduction}
\label{sec:cc1}
Binary stars are common in the Galaxy \citep[e.g.,][]{1991A&A...248..485D}. 
They deserve to be considered when studying stellar kinematics, star formation rate, initial mass function 
and the occurrence of special stars, as aspects of the study of stellar populations.
Given the light curves or radial velocity (RV)  variations with time, binary stars provide an independent 
method to obtain stellar masses and radii using Kepler's third law, which offers exciting opportunities to 
develop highly constrained stellar models  \citep[e.g.,][]{2014MNRAS.441.2316G,2016ApJ...832..121G,2016ApJS..227...29P}. 
Material and energy exchanges between the members of binary system create the ideal breeding ground 
for the formations of some special stars, such as type Ia supernova. 
Binary interactions make the appearance of the population substantially bluer, which affect the derived 
age and metallicity of the population \citep{2007IAUS..241..205Z}. 
Moreover, physical processes unique hosted to close-in binary systems such as mass exchange, 
are far from well understood. 

It's important to identify binary stars or estimate binary fractions as a precondition to study properties of these stars. 
Several methods have been developed to detect binary stars, such as by means of radial velocity, astrometric acceleration, direct imaging, and common proper motion. 
Based on a complete sample of dwarf and subdwarf stars from the \textit{Hipparcos} catalog \citep{1997ESASP1200.....E},  \citet{2010ApJS..190....1R}  analyze the multiplicity of solar-type stars out to 25 pc. 
They find that the massive stars are more likely to have companions than less massive ones. Meanwhile, statistical analyses of quantities such as binary period distribution, mass-ratio distribution for binary systems are presented by \citet{2010ApJS..190....1R}.
Through the light curves obtained with the \textit{Kepler} \citep{2010Sci...327..977B,2010ApJ...713L..79K}, 
 a database\footnote[1]{Kepler Eclipsing Binary Catalog: \\ http://keplerebs.villanova.edu} of 2876 eclipsing binaries (updated on Apr. 27, 2017) has been constructed, including five groups: detached, semi-detached, over-contact, ellipsoidal binaries, and uncertain \citep{Mat12}. 
The probability of a star having a companion could be determined through comparing observed RVs with a single-star model and a binary-star model using Markov chain Monte Carlo (MCMC) method \citep{2015ApJ...806L...2H}, and metal-rich disk stars are found to be 30\% more likely to have companions with periods shorter than 12 days than metal-poor halo stars. However, this method probably underestimates the fraction of binary stars. 
Rather than estimating the probability of a star having a companion, the total binary star fraction could be statistically measured by the dispersion of RVs from multi-epoch observations \citep{2012ApJ...751..143M,2014ApJ...788L..37G,2017MNRAS.469L..68G}. 
Single stars and binary stars have different distributions on color-color diagram, thus, binary fraction of main-sequence (MS) stars can also be estimated statistically through a stellar locus outlier (SLOT) method \citep{2015ApJ...799..135Y}.
The previous work by \citet{2014ApJ...788L..37G,2017MNRAS.469L..68G} and \citet{2015ApJ...799..135Y} show that metal-poor stars are more likely to possess companions than metal-rich ones. 
An analysis by  \citet{2017arXiv171100660B} based on the APOGEE data \citep{2017AJ....154...94M} suggests that the metal-poor stars have a multiplicity fraction a factor 2-3 higher than metal-rich stars, which is in qualitative agreement with the works of \citet{2014ApJ...788L..37G,2017MNRAS.469L..68G}  using low-resolution spectra of main sequence stars. 

The multiplicity of F and G-type dwarf stars within 67 pc of the Sun with a completeness greater than 90\% are presented by \citet{2014AJ....147...86T}, and 80\% of companions to the main targets in the sample are detected, which shows that the multiplicity fraction of the sample is about 46\%.  
\citet[][]{2017arXiv170607095S} estimates the mean binary fraction of extreme young stars from young moving groups (YMGs) about $35^{+5}_{-4}$ \%, which is lower than that of F and G-type field stars.  
Meanwhile, a combined analysis of multiplicity fraction as a function of mass and age over the range of 0.2 to 1.2 $\rm M_{\odot}$ and 10 to 200 Myr appears to be linearly flat in both parameters and across YMGs, which suggests constant multiplicity fractions from YMGs despite their differences in age and possibly birth environments \citep[][]{2017arXiv170607095S}. 
The binary fraction of white dwarf systems ($\sim25\%$) within 20 pc is much lower compared to solar-type stars ($\sim50\%$), which is explained by mergers in binary systems \citep[][]{2017A&A...602A..16T}.  
Research of local white dwarf population within 25 pc suggests mechanisms that result in the loss of companions during binary system evolution \citep[][]{2016MNRAS.462.2295H}.  
The observed binary fraction is the combination of birth and dissipation rates of binary systems. 
In order to understand the observed binary fraction well, we'd better to estimate binary fractions of stars with different ages and evolution stages.

With the increase in quantity of data, more researches about binary stars could be carried out. 
In Section \ref{sec:chap2}, we describe the method and assumptions adopted in this work. The data is presented in Section \ref{sec:chapdata}. The results are shown in section \ref{sec:res}, followed by discussions and conclusions in section \ref{sec:cd}.

\section{Method of Measuring Binary star fraction}
\label{sec:chap2}
Binary systems can be detected through the dispersion of radial velocities  ($\Delta \rm{RV}$) obtained with multi-epoch observations for the same stars. 
Through analyzing the distribution of  the maximum radial velocity difference ($\Delta \rm{RV_{max}}$) between any two epochs for the same object, \citet{2012ApJ...751..143M} statistically characterized binary fraction of a detached binary sample observed  with the SDSS \citep{2000AJ....120.1579Y}. 
The core of the $\Delta \rm{RV_{max}}$ distribution is dominated by measurement errors, and the tail of the distribution is considered as a contribution from the binary population. 

Although low-resolution spectra could not provide RV with high-enough precision to identify binarity for every star, we could statistically estimate binary fraction of a sample through a database of RV from low-resolution spectra. 
Since the measurement of a quantity equals to the true value plus the observational error, the $\Delta \rm{RV}$ for single stars are determined by  the measurement uncertainty, while these for binary stars are determined by the phase of binary system and the uncertainty of RVs. Thus, the distributions of $\Delta \rm{RV}$ would be described as \citep[][]{2014ApJ...788L..37G}:
\begin{equation}
   p(\Delta \rm{RV} )= \textit{f}_{B} \,  \textit{p}_{B}(\Delta \rm{\rm RV}|\sigma _{\rm RV}, \Delta \mathit{\rm t}, M_{B}) + (1- \textit{f}_{B} )\,  \textit{p}_{S}(\Delta \rm{RV}|\sigma _{\rm RV}).
	\label{eq:f1}
\end{equation}
where $\textit{f}_{\rm B}$ and $\sigma _{\rm RV}$ denote binary fraction and radial velocity uncertainty, respectively, and $\Delta \rm t$ represents time interval between two observations. 
$ \textit{p}_{\rm B}(\Delta \rm{RV}|\sigma _{\rm RV},\Delta \rm t, M_{B}) $ and $\textit{p}_{\rm S}(\Delta \rm{RV}|\sigma _{\rm RV})$ denote probabilities of obtaining $\Delta \rm{RV}$ under the assumptions of binary stars and single stars, respectively. 
Note that here we ignore the fraction of multiples (number of objects in the system $\geq $3), which is much lower than that of single or binary stars \citep{2010ApJS..190....1R,2014AJ....147...87T}. 

For the binary systems $\rm M_{B}$, we suppose that : (1) the observed RVs are contributed from the primary stars of the binary systems; 
(2) their orbital orientations are considered as isotropic in 3D space and initial phases follow uniform distributions. 
Meanwhile, the method adopted in this work is sensitive to orbital period distribution and mass ratio distribution of binary systems.
In order to test effects of these distributions of binary stars on the determinations of $\textit{f}_{\rm B}$ and $\sigma _{\rm RV}$, in this paper, we consider 8 possible permutations of two models for the mass distribution of primary stars (Salpeter IMF $\&$ the mass distribution derived from isochrones),  
two models for the mass ratio distribution (a power-law distribution $f(q) \propto q^{\gamma}$ $\&$ a uniform distribution) 
and two models for the orbital period distribution ($\log P \sim N(5.03,2.28^{2})$  $\&$ $f(P) \propto P^{\alpha}$). 
An overview of these  models is given in Table \ref{tab:example_table1} and their details are explained below.

\begin{figure}
\centering\includegraphics[width=0.7\columnwidth]{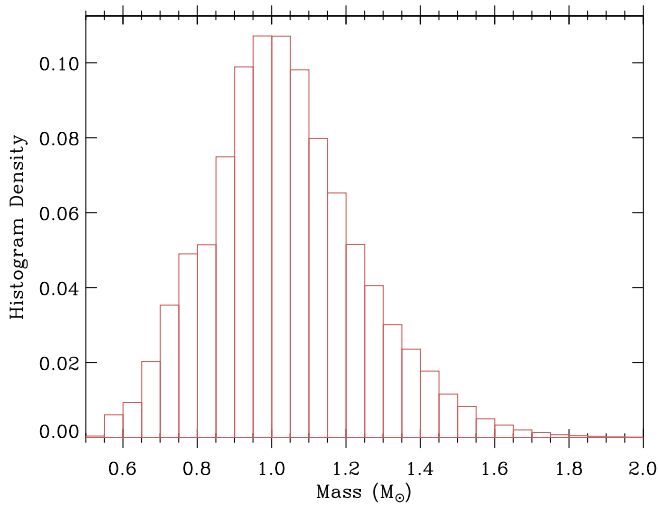}
\caption{Mass distribution of stars adopted in this work.} 
\label{fig:mdis}
\end{figure}

Two different mass distributions of primary stars are adopted in this work, one following the Salpeter IMF \citep[][]{1955ApJ...121..161S}, in which the masses of stars obey a power-law distribution, the other following the measured mass distribution of observed stars. 
Their masses and other fundamental parameters are determined by comparing the atmospheric parameters with the YY isochrones \citep[][and references therein]{2004ApJS..155..667D}. 
The details of estimating fundamental parameters are described in section \ref{sec:res43}. 
As presented in Fig. \ref{fig:mdis}, the mass distribution of the whole sample shows a peak around 1 $\rm M_{\odot}$ since the sample is not complete. 
For the mass ratio distributions, a power-law model $f(q) \propto q^{\gamma}$ \citep[e.g.][where $\gamma =0.3 \pm 0.1$ for F, G and K-type main sequence stars]{2010ApJS..190....1R} and a uniform distribution are adopted. 
For the orbital period distributions, a log-normal Gaussian profile \citep[with a mean value of $\log P$ = 5.03 and a dispersion value of $\sigma_{\log P}$ = 2.28, where $P$ is in units of days, see ][]{2010ApJS..190....1R} and a power-law distribution function $f(P) \propto P^{-1}$ \citep[][]{1924PTarO..25f...1O,1983ARA&A..21..343A} are adopted for comparisons. 
Note that the distributions of periods and mass ratios in \citet{2010ApJS..190....1R} are derived from a highly complete sample of solar-type stars within 25 pc. 

\begin{table*}
	\centering
	\caption{The assumptions of period distribution, primary mass and mass ratio.}
	\label{tab:example_table1}
	\begin{tabular}{lccr} 
	\hline
	Model No. & Orbital period distribution & Primary mass distribution & Mass ratio distribution \\
	\hline
	1  &  $\log P \sim N(5.03,2.28^{2})$ \color{red}$^a$  & $\xi (\rm M)=\xi_{0}{\rm M}^{-2.35}$  \color{red}$  ^c$   & $f(q) \propto q^{0.3 \pm 0.1}$ \color{red}{$^a$}   \\
	2  &  $\log P \sim N(5.03,2.28^{2})$ \color{red}$^a$  & $\xi (\rm M)=\xi_{0}{\rm M}^{-2.35}$  \color{red}$  ^c$  & uniform distribution              \\
	3  &  $\log P \sim N(5.03,2.28^{2})$ \color{red}$^a$  &  masses from the sample  \color{red}{$^d$} & $f(q) \propto q^{0.3 \pm 0.1}$  \color{red}{$^a$}  \\
	4  &  $\log P \sim N(5.03,2.28^{2})$ \color{red}$^a$  & masses from the sample  \color{red}{$^d$} & uniform distribution              \\
        5  &  $f(P) \propto P^{-1}$  \color{red}{$^b$} &  $\xi (\rm M)=\xi_{0}{\rm M}^{-2.35}$  \color{red}$  ^c$    & $f(q) \propto q^{0.3 \pm 0.1}$  \color{red}{$^a$}   \\
        6  &  $f(P) \propto P^{-1}$ \color{red}{$^b$}   &  $\xi (\rm M)=\xi_{0}{\rm M}^{-2.35}$  \color{red}$  ^c$    & uniform distribution               \\
        7  &  $f(P) \propto P^{-1}$  \color{red}{$^b$}   & masses of the sample  \color{red}{$^d$}  & $f(q) \propto q^{0.3 \pm 0.1}$ \color{red}{$^a$}   \\
        8  &  $f(P) \propto P^{-1}$   \color{red}{$^b$}  & masses of the sample  \color{red}{$^d$}  & uniform distribution  \\
	\hline
\multicolumn{4}{l}{\color{red} $^a$ \color{black}  \citet[][]{2010ApJS..190....1R}; }\\
\multicolumn{4}{l}{\color{red} $^b$ \color{black}  \citet[][]{1924PTarO..25f...1O,1983ARA&A..21..343A};}\\
\multicolumn{4}{l} {\color{red} $^c$ \color{black} \citet[][]{1955ApJ...121..161S}; }\\
\multicolumn{4}{l}{\color{red} $^d$ \color{black}  masses of the sample derived from theoretical isochrones.}\\
	\end{tabular}
\end{table*}

The time separations of multi-epoch observations in this work (with a typical value less than 5 yrs) are much shorter than the mean period of binary stars \citep[with a typical period of about 300 yrs, see][]{2010ApJS..190....1R}. 
The method adopted in this work is more sensitive to estimate the binary fraction of stars with short periods than long periods. 
While short-period binaries probably dominate circular or sub-orbicular orbits under the action of gravity.
 Thus, we ignore the distribution of eccentricities and assume the circular orbits for binary systems in our study.

The distribution of $\Delta \mathit{\rm t}$ used in the calculation is that of the stars considered in each bin. 
Given the distribution of $\Delta \mathit{\rm t}$, the distribution of $\Delta \rm{RV}$ would be a function of $\textit{f}_{\rm B}$ and $\sigma _{\rm RV}$.
Through the distribution, $p(\Delta \rm{RV} )$, we could estimate the binary fraction $\textit{f}_{\rm B}$ and the mean error of radial velocities $\sigma _{\rm RV}$, with a maximum likelihood estimate method (MLE).
The uncertainties of radial velocities are functions of signal noise ratios (SNR), $T_{\rm eff}$, $\log g$ and [Fe/H] \citep{2015MNRAS.448..822X}.  
For single stars with similar SNR, $T_{\rm eff}$, $\log g$ and [Fe/H], the distribution of $\Delta \rm{RV}$ follows a Gaussian distribution. However, the distribution of $\Delta \rm{RV}$ for binary stars doesn't follow a Gaussian distribution, because the RV varies with time or phase of binary system. The wings of the distribution overleaping the Gaussian profile are engendered from the contribution of binary stars. Thus, we could estimate the fraction of binary stars through the distribution of $\Delta \rm{RV}$. 
However, the combinations of $\Delta \rm{RV}$ with different dispersions could also deduce wings overleaping a Gaussian profile. 
There is some degeneracy between $\textit{f}_{\rm B}$ and $\sigma _{\rm RV}$. To minimize the effects of $\sigma _{\rm RV}$ dispersions on the estimation of  $\textit{f}_{\rm B}$, it's necessary to construct subsamples with similar $\sigma _{\rm RV}$. 
Therefore, only data with SNR greater than 50 are used in the analyses. 
Meanwhile, we estimate binary star fractions in multi-dimensional bins of $T_{\rm eff}$, [$\alpha$/Fe] and [Fe/H]. 
Note that the primary mass distributions in each bin differ because of the restrictions of $T_{\rm eff}$, [$\alpha$/Fe] and [Fe/H]. 
To ensure the validity of estimation, we only consider the bins with star numbers over 1500.

\section{Data}
\label{sec:chapdata} 
With the progress of Large Sky Area Multi-Object Fiber Spectroscopic Telescope (LAMOST) spectroscopic survey \citep{2012RAA....12.1197C,2012RAA....12..723Z,2012RAA....12..735D,2014IAUS..298..310L,2015MNRAS.448..855Y}, more and more stars are observed more than once, providing a great opportunity for studies of field binaries. 
About 4 millions unique objects marked with `star' have been targeted by Nov, 2016, 
and over 0.93 million unique stars with SNR greater than 10 have been observed with two or more epochs. 
To obtain atmospheric parameters and radial velocities of stars, stellar parameter pipelines such as the LASP \citep{2011RAA....11..924W} and the LSP3 \citep{2015MNRAS.448..822X,2017MNRAS.464.3657X} have been developed and applied to the LAMOST spectra. 
We adopt the stellar parameters yield by the LSP3 for F, G and K-type stars observed with the LAMOST (by November, 2016).

\begin{figure}
\centering\includegraphics[width=0.7\columnwidth]{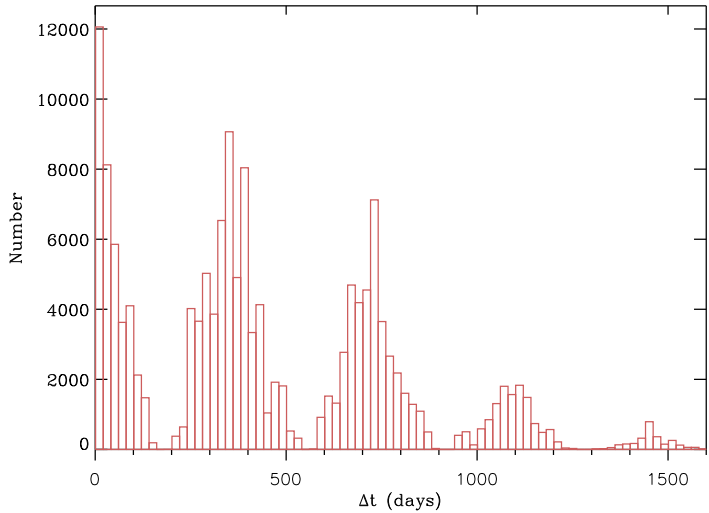}
\caption{Time interval distribution of 0.15 million stars used in this work.} 
\label{fig:dt_copy}
\end{figure}

We focus on the binary fractions of F, G and K-type main-sequence stars. In order to remove giant stars from the sample, 
a cut of  $\log g>$ 3.75 is adopted.
The $T_{\rm eff}$ of selected sample is limited in the range of 4000 - 7000 K. 
The observations at the same night probably share similar conditions (e.g. seeing, skylight, flat), 
and the measured velocities share similar systematic errors. 
Therefore, the data with time intervals less than one day is excluded in this work to make sure that multi-epoch observations for each star were carried out in different nights. 
For most binary stars, the periods are about dozens to hundreds of years. 
Thus, the method is more sensitive to multi-epoch observations with long time intervals than short ones.  
In principle, RVs with high SNR and large time separation of multi-epoch observations are required to carry out the work. 
For stars observed more than twice, we obtain a vector of radial velocity $\rm RV=\{ \rm RV_{1}, \rm RV_{2}, ... \rm RV_{n}\}$ response for a vector of time $\rm t=\{ \rm t_{1}, \rm t_{2}...\rm t_{n}\}$, we select the two RVs which dominate the maximum value of $\rm SNR_{j}*\rm SNR_{k}* (\rm t_{j}- \rm t_{k})$. 
We obtain about 0.15 million stars observed at least two times and with time interval larger than one day and SNR greater than 50. 
The time interval distribution of multi-epoch observations for the sample is presented in Fig. \ref{fig:dt_copy}.  The maximum of time intervals is about 1600 days. 

\section{Binary fractions in parameter spaces}
\label{sec:res}

\begin{figure}
\centering\includegraphics[width=0.52\textwidth]{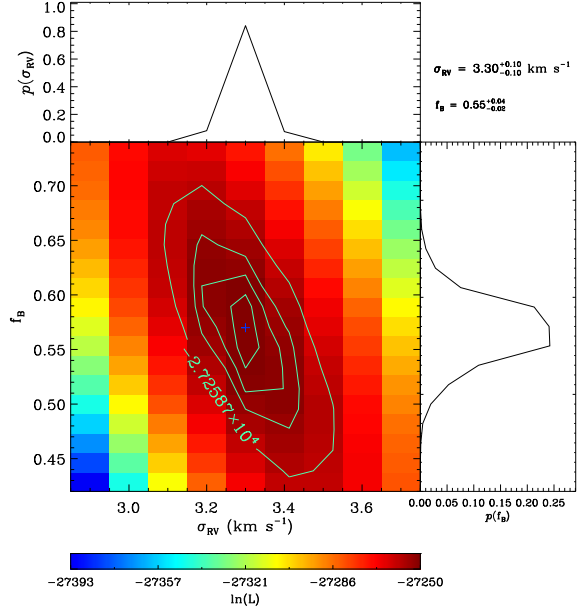} 
\caption{Likelihood of $\sigma _{\rm RV}$ and $\textit{f}_{\rm B}$ for stars with $T_{\rm eff}$ in the range of 5000 - 5500 K,  [Fe/H] in the range of $-$0.10 - 0.10 dex, and [$\alpha$/Fe] in the range of $-$0.10 - 0.10 dex. The top and right panels present the probabilities of $\sigma _{\rm RV}$ and $\textit{f}_{\rm B}$, respectively. The values of $\sigma _{\rm RV}$ and $\textit{f}_{\rm B}$ with 1-sigma error are shown on the up-right.} 
\label{fig:mle}
\end{figure}

\begin{figure}
\centering\includegraphics[width=0.52\textwidth]{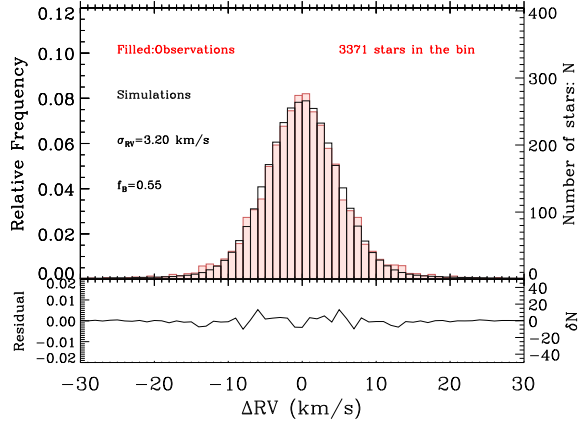}
\caption{The top panel shows the observed (red-filled) and fitted (black-unfilled) $\Delta \rm RV$ distribution for solar-type stars. 
The residuals are plotted on the bottom panel.} 
\label{fig:fit}
\end{figure}

\begin{figure*}
\centering\includegraphics[width=0.8\textwidth,angle=0]{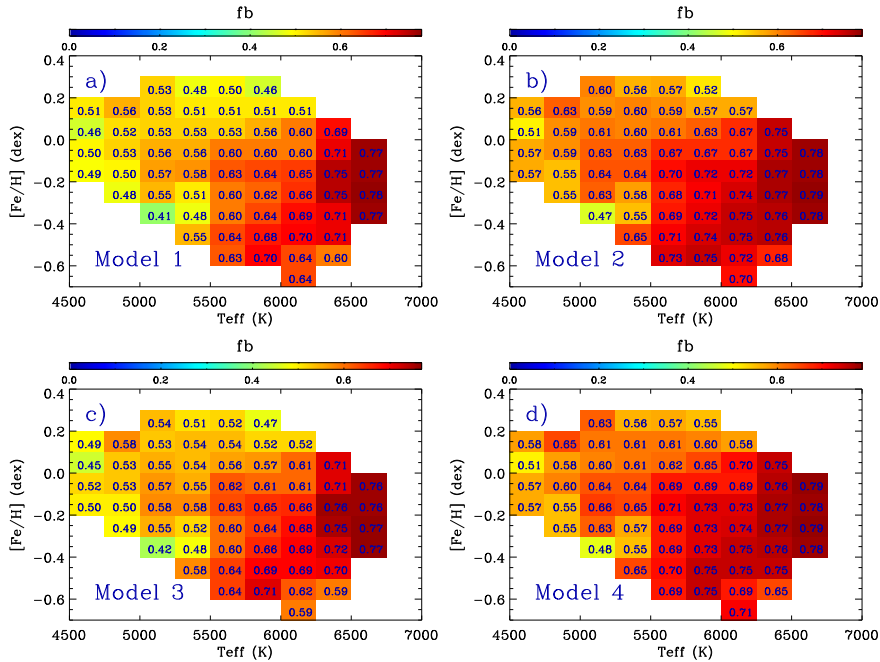}
\caption{Binary fractions of stars on the $T_{\rm eff}$-[Fe/H] panel under different model assumptions. The colors denote the values of binary fractions in different bins. 
The errors of $\textit{f}_{\rm B}$  are below 10\%.}
\label{fig:model1}
\end{figure*}

\begin{figure*}
\centering\includegraphics[width=0.8\textwidth,angle=0]{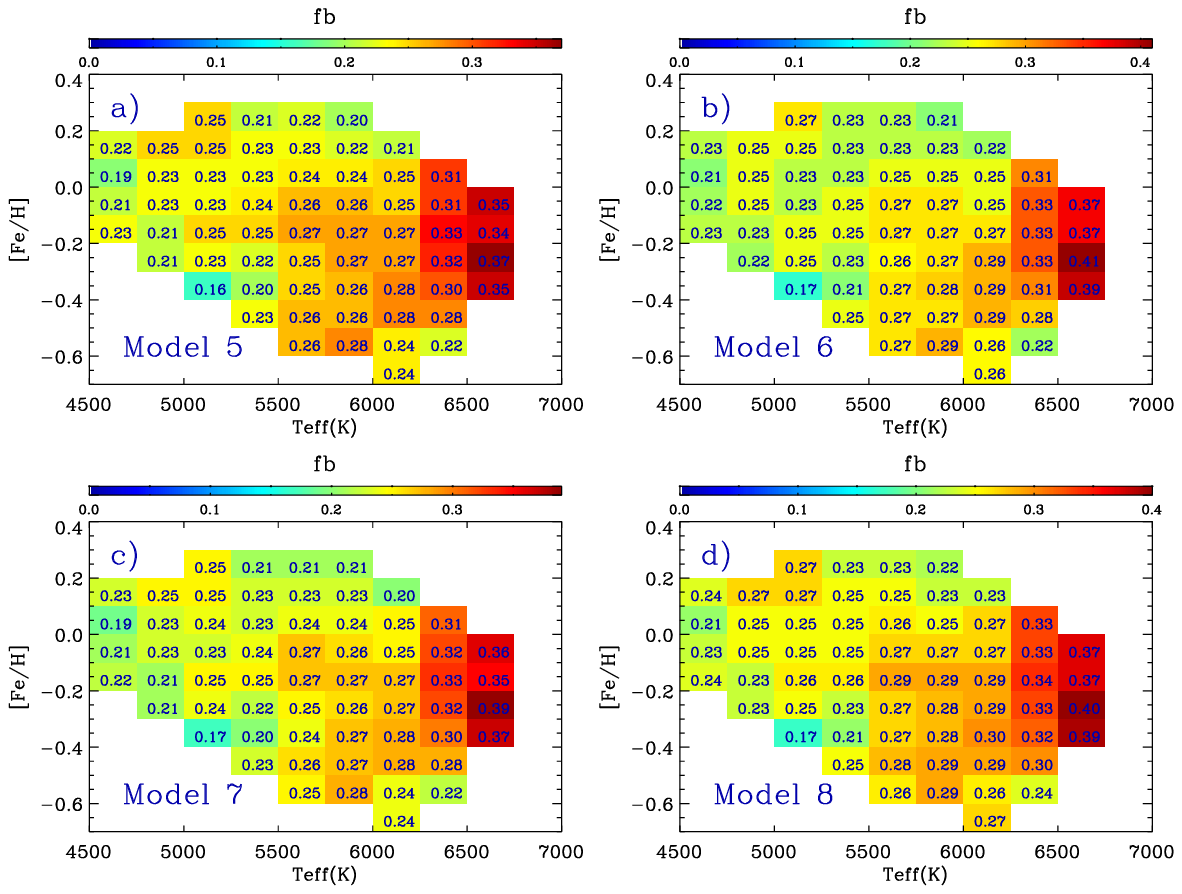}
\caption{Same as Fig. \ref{fig:model1}, but for Models 5-8.}
\label{fig:model7}
\end{figure*} 

The sample is divided into subsamples with 3-dimensional bins of $T_{\rm eff}$, [Fe/H] and [$\alpha$/Fe]. The binning width of these quantities are 500 K, 0.2 dex and 0.2 dex with overlaps of 250 K, 0.1 dex and 0.1 dex, respectively. 
The binary fractions are estimated for each subsample and under different model assumptions listed in Table 1. 
For example, Fig. \ref{fig:mle} shows the contour of the likelihood of $\sigma _{\rm RV}$ and $\textit{f}_{\rm B}$ for stars with $T_{\rm eff}$ in the range of 5000 - 5500 K,  [Fe/H] in the range of $-$0.10 - 0.10 dex, and [$\alpha$/Fe] in the range of $-$0.10 - 0.10 dex. 
The estimation is performed under the assumptions of Model 1. 
The two-dimensional joint probability is converted into the probabilities of $\sigma _{\rm RV}$ and $\textit{f}_{\rm B}$, namely $p(\sigma _{\rm RV})$ and $p(\textit{f}_{\rm B})$, respectively. 
The values of $\sigma _{\rm RV}$ and $\textit{f}_{\rm B}$ and their 1-sigma errors are estimated accordingly. 
Our result, $\textit{f}_{\rm B}=55^{+4}_{-2}  \%$, is consistent  with previous work \citep[e.g.][]{2010ApJS..190....1R}. 
A comparison between observed and simulated $\Delta \rm RV$ distributions is plotted on Fig. \ref{fig:fit}. As shown on the bottom panel of the figure, the residual is bellow 1\%. 

\subsection{The effects of different assumptions on the estimation of binary fractions}
In order to compare the results under different assumptions, we plot the binary fractions against $T_{\rm eff}$ and [Fe/H] for stars of $-$0.1 $<$  [$\alpha$/Fe] $<$ 0.1 
based on Models 1 - 4 on Fig. \ref{fig:model1} and Models 5 - 8 on Fig. \ref{fig:model7}, respectively. 
All models on Fig. \ref{fig:model1} are based on the log-normal distribution of orbital periods \citep[][]{2010ApJS..190....1R},
but assuming different mass distributions of the primary stars and mass ratio distributions.

For Models 1 - 4, the discrepancies of $\textit{f}_{\rm B}$ are below 10 percent. 
For Models 5 - 8, the estimated $\textit{f}_{\rm B}$ values based on the power-law distribution of orbital period are about half of that based on the log-normal Gaussian distribution. 
This is because that binary systems of Models 5 - 8 have a larger fraction of short-period binaries than Models 1 - 4. 
The statistical analysis of the complete sample within 25 pc suggests that binary fraction of solar-type stars is about 50\% \citep[][]{2010ApJS..190....1R}, which supports the result based on the log-normal Gaussian distribution of orbital periods. 
In the following analyses, we only consider the results based on Models 1 - 4. 

Given the orbital periods following the log-normal distribution, $\textit{f}_{\rm B}$ values estimated based on different primary mass distributions and mass ratio distributions differ slightly. 
Models with the same orbital period distribution and the same primary mass distribution as Models 1 and 2 (or Models 3 and 4) show that the $\textit{f}_{\rm B}$ estimated under the mass ratio distribution of a power-law index is higher than that based on a uniform distribution. 
This is because that given the period and mass ratio distributions, binary systems with high primary mass will obtain a large separation and a high velocity. 

Comparing Models 1 and 3 (or Models 2 and 4), we find that the $\textit{f}_{\rm B}$ estimated under the primary mass distribution of the \citet{1955ApJ...121..161S}  IMF is slightly lower than that based on the mass distribution of the sample. 
The order of mean $\textit{f}_{\rm B}$ from large to small is Models 4, 2,  3 and 1. 
Even the difference of $\textit{f}_{\rm B}$ between Models 1 and 4 is lower than 10 percent. 

\subsection{Binary fractions on the $T_{\rm eff}$-[Fe/H] panel}
\label{sec:res42}

As shown in all panels of Fig. \ref{fig:model1}, for a given value of [Fe/H], the binary fraction increases as the $T_{\rm eff}$ increases for main sequence stars.
For a given value of $T_{\rm eff}$, the binary fraction decreases as the  [Fe/H] increases. 
There is a positive gradient of binary fractions from the top left to the bottom right on the $T_{\rm eff}$-[Fe/H] panel. 
The results agree well with that of \citet[][]{2017MNRAS.469L..68G}. 

\begin{figure*}
\centering\includegraphics[width=0.78\textwidth]{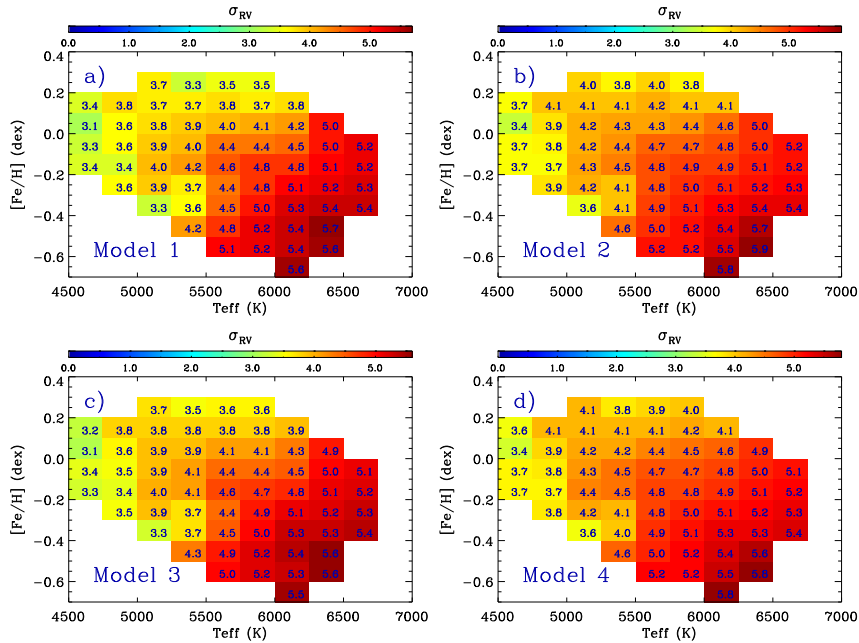}
\caption{Uncertainties of  radial velocity against the $T_{\rm eff}$-[Fe/H]. The panels a - d present the $\sigma _{\rm RV}$ estimated based on models 1 - 4 on Fig. \ref{fig:model1}, respectively. The colors denote the values of $\sigma _{\rm RV}$  in different bins.}
\label{modelsig}
\end{figure*}

As described in \citet[][]{2015MNRAS.448..822X}, the $\sigma _{\rm RV}$ variation is a function of SNR,  $T_{\rm eff}$ and [Fe/H].  
In Fig. \ref{modelsig}, we present the distribution of $\sigma _{\rm RV}$ on the $T_{\rm eff}$-[Fe/H] panel corresponding to the $\textit{f}_{\rm B}$ in Fig. \ref{fig:model1}. 
We find a positive correlation coefficient between $\sigma _{\rm RV}$ and $\textit{f}_{\rm B}$ about 0.70. 
There is a strong negative correlation between $\sigma _{\rm RV}$ and $\textit{f}_{\rm B}$,  as shown in Fig. \ref{fig:mle}, suggesting the trend of $\textit{f}_{\rm B}$ against $T_{\rm eff}$ and [Fe/H] is not caused by the trend of $\sigma _{\rm RV}$ against stellar parameters. 

\subsection{Binary fractions against star formation time}
\label{sec:res43}

\begin{figure*}
\centering\includegraphics[width=0.8\textwidth,angle=0]{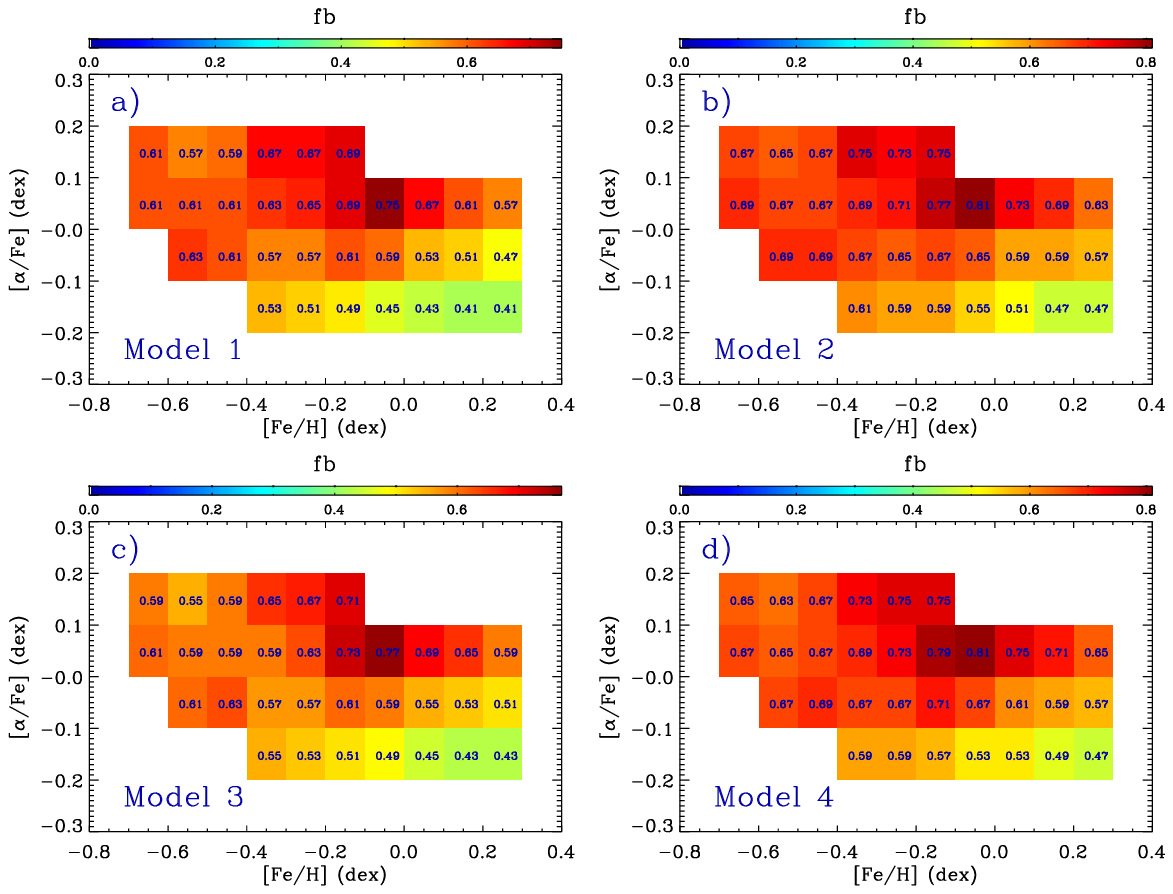} 
\caption{Binary fractions of stars with $T_{\rm eff}$ in the range of 5500 - 6000 K on the [Fe/H]-[$\alpha$/Fe] panel. The colors denote the values of binary fractions in different bins.}
\label{fig:hist2dinfobin5500_6000_feh_afe_dr4}
\end{figure*}

\begin{figure*}
\centering\includegraphics[width=0.80\textwidth]{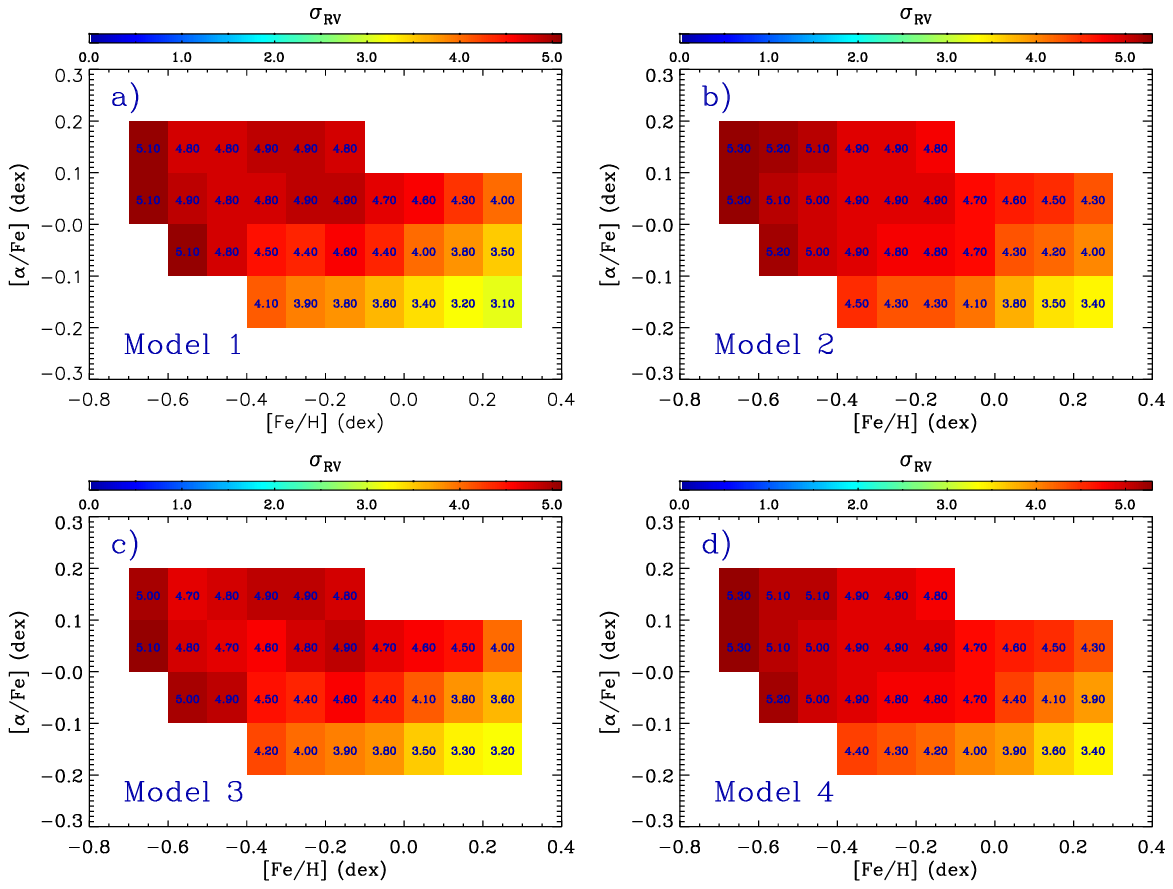}
\caption{The uncertainty of radial velocity against the [Fe/H]-[$\alpha$/Fe]. The panels a - d present the $\sigma _{\rm RV}$ estimated based on models 1 - 4 on Fig. \ref{fig:hist2dinfobin5500_6000_feh_afe_dr4}, respectively. The colors denote the values of $\sigma _{\rm RV}$  in different bins.}
\label{modelsigafe}
\end{figure*}

According to the theory of stellar evolution, high-mass stars evolve faster than low-mass ones during the main-sequence phase. The typical age of high-mass MS stars is younger than low-mass ones. 
The phenomena that binary fractions of high-$T_{\rm eff}$  stars are higher than these of low-$T_{\rm eff}$ main sequence stars is probably related with the variation of $\textit{f}_{\rm B}$ against masses or formation times of stars. 
Given a star's current age, we can trace the time of its formation. The older the star is, the earlier it formed.
To study the binary fractions in stellar parameter spaces, it's necessary to obtain the fundamental parameters such as mass and formation time of stars. 
For each main-sequence star, we start by using [Fe/H], [$\alpha$/Fe], $T_{\rm eff}$ and $\log g$ to get an estimate of the mass (M) and formation time (t) of the star using the Yale Birmingham Grid-based modeling pipeline \citep[][]{2010ApJ...710.1596B}, which estimates stellar properties by comparing observed quantities to the outputs of stellar evolutionary models. 
The grid-based stellar models adopted in this work are from $Y^2$ isochrones \citep[][and reference therein]{2004ApJS..155..667D}, which consider the effects of [$\alpha$/Fe]. 

Figure \ref{fig:hist2dinfobin5500_6000_feh_afe_dr4} presents the binary fraction variations with the metallicities and $\alpha$-element abundances for stars with $T_{\rm eff}$ in the range of 5500 - 6000 K. 
The $\sigma _{\rm RV}$ variations corresponding to models 1 - 4 on Fig. \ref{fig:hist2dinfobin5500_6000_feh_afe_dr4} are presented on Fig. \ref{modelsigafe}. 
For $\alpha$-poor stars, it shows the trend that binary fractions increase with the decrease of [Fe/H], which is coincident with  Fig. \ref{fig:model1}. 
For $\alpha$-rich stars, the binary fractions show obvious variations with [Fe/H]. 
There's a macroscopical trend that the $\alpha$-rich stars possess higher $\textit{f}_{\rm B}$ than $\alpha$-poor stars. 
Since the elements abundances are indicators of star formation time, the variations of binary fractions against elements abundances reflect the variation of binary fractions with time, which may be related to the evolution of the Galaxy.

\begin{figure*}
\centering\includegraphics[width=0.85\textwidth,angle=0]{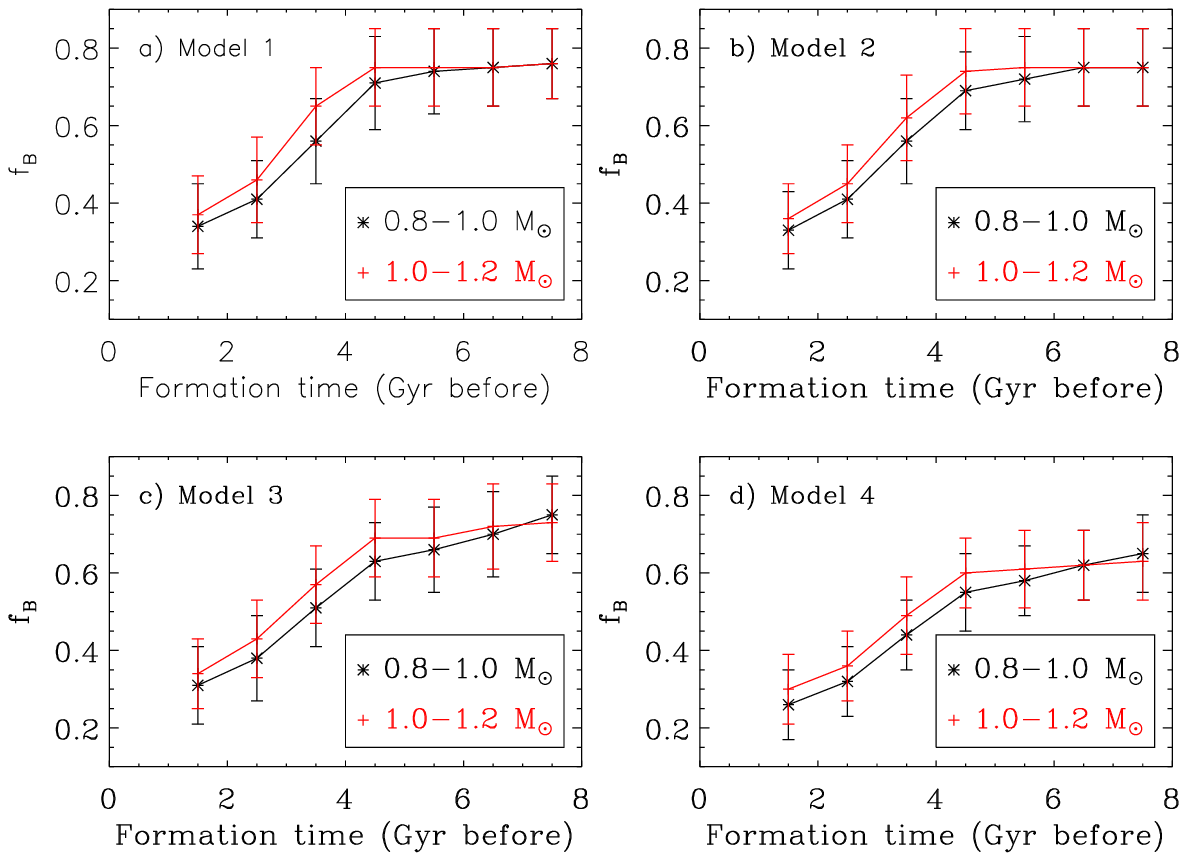}
\caption{Binary fraction as a function of star formation time. The range of metallicity is $-$0.1 - 0.1 dex. }
\label{fig:modela}
\end{figure*}

In order to test the changes of binary fraction with star formation time directly, we estimate the binary fractions in two-dimensional bins of mass and star formation time. 
We select samples of low-mass stars formed 1 - 8 Gyr ago with [Fe/H] in the range of $-$0.1 - 0.1 dex. 
The selected stars between 0.8 and 1.2 $\rm M_{\odot}$ are divided into two bins of equal width of mass. 
The relative errors of mass and formation time are about below 10\% and 50\% for this sample, respectively. 
Figure \ref{fig:modela} presents the binary fraction as a function of star formation time. 
It shows that the former forming stars possess a higher binary fraction than newly forming ones.
Given the mass difference of 0.2 $\rm M_{\odot}$, 
the binary fraction of high-mass stars is higher than that of low-mass stars. 
Note that the star formation rates (SFR) based on integrated-light analyses and stellar color-magnitude diagrams show good agreement that SFR is a function of time \citep[][]{2015A&A...583A..60R}. More researches should be carried out to study the relation between the initial binary fraction and SFR. 

\section{Conclusions and Discussions}
\label{sec:cd}
In this paper, we estimate the binary fractions $\textit{f}_{\rm B}$ of 0.15 million dwarf stars observed with the LAMOST. 
The estimated $\textit{f}_{\rm B}$ of the sampled stars are sensitive  to the adopted assumptions of mass function, mass ratio distribution and period distribution.  
The estimated $\textit{f}_{\rm B}$ about 50\% for solar-type stars based on log-normal distribution of orbital periods is coincident with previous statistical analysis of a complete sample within 25pc \citep[e.g.][]{2010ApJS..190....1R}. 
Not only in small sample, but also in such a large sample of survey data the orbital periods of binary stars prefer a log-normal distribution to other ones. 

The binary fractions increase with the increasing of  $T_{\rm eff}$ and the decreasing of [Fe/H]. 
We first investigate the relation between $\alpha$-elements and binary fraction in such a large sample as the LAMOST. 
The old stars with high [$\alpha$/Fe]  have a higher binary fraction than young stars with low [$\alpha$/Fe]. 
The variations of binary fraction with star formation time may be related with the evolution of the Galaxy. 

This work is based on the assumption that RVs of binary stars are derived from the spectra of primary stars. 
The binaries with similar components are difficult to detect, since their spectra are blended. However, this bias doesn't change the trend of binary fractions in space bins. The effects of spectra blend on the estimating of binary fractions will be discussed in our future work. 

With the progresses of sky-surveys such as SDSS \citep{2000AJ....120.1579Y} and LAMOST  \citep{2012RAA....12.1197C,2012RAA....12..723Z}, RVs of more and more stars could be derived from spectra. 
Radial velocities from spectra surveys together with parallaxes from Gaia \citep{2016A&A...595A...1G} make that researches about binary stars are not limited to a small sample of data. 
Large sample analysis of binary stars will improve our knowledge about binary star formation and dissipation all over the Milky way.  
In the future work, we plan to identify binary stars through their RV and proper motion variations. From the LAMOST and Gaia data of millions of stars, a huge gallery of identified binary stars could be established. Enormous number of binary stars will be used to trace the evolution of the Galactic stellar populations. 

\normalem
\begin{acknowledgements}
This work has made use of data products from the Guoshoujing Telescope (the Large Sky Area Multi-Object Fibre Spectroscopic Telescope, LAMOST). LAMOST is a National Major Scientific Project built by the Chinese Academy of Sciences. Funding for the project has been provided by the National Development and Reform Commission. LAMOST is operated and managed by the National Astronomical Observatories, Chinese Academy of Sciences.

This work is partially supported by National Key Basic Research Program of China 2014CB845700, China Postdoctoral Science Foundation 2016M600850,
National Natural Science Foundation of China 11443006 and Joint Research Fund in Astronomy U1531244 and U1631236. 
The LAMOST FELLOWSHIP is supported by Special Funding for Advanced Users, budgeted and administrated by Center for Astronomical Mega-Science, Chinese Academy of Sciences (CAMS).
\end{acknowledgements}
  
\bibliographystyle{raa}
\bibliography{example}

\end{document}